\documentclass[prc, showpacs, showkeys, superscriptaddress, preprint, onecolumn]{revtex4}
\usepackage{graphicx}
\usepackage[bookmarksopen, bookmarksnumbered]{hyperref}

\newcommand{\mevu}{\mbox{MeV/nucleon}}

\newcommand{\nscl}{\affiliation{National Superconducting Cyclotron Laboratory and Department of Physics \& Astronomy, Michigan State University, East Lansing, MI 48824, USA}}

\newcommand{\caen}{\affiliation{Ganil, CEA, IN2P3-CNRS, B. P. 5027, F-14021 Caen Cedex, France}}
\newcommand{\tohoku}{\affiliation{Department of Physics, Tohoku University, Sendai 980-8578, Japan}}
\newcommand{\wu}{\affiliation{Chemistry Department, Washington University, St. Louis, MO 63130, USA}}
\newcommand{\lanl}{\affiliation{Los Alamos National Laboratory, P. O. Box 1663, Los Alamos, NM 87545, USA}}

\begin{document}
\title{Transport Model Simulations of Projectile Fragmentation Reactions at 140 MeV/nucleon}
\author{M.~Mocko}\lanl\nscl
\author{M.~B.~Tsang}\nscl
\author{D.~Lacroix}\nscl\caen
\author{A.~Ono}\tohoku
\author{P.~Danielewicz}\nscl
\author{W.~G.~Lynch}\nscl
\author{R.~J.~Charity}\wu

\begin{abstract}
The collisions in four different reaction systems using $^{40,48}$Ca and $^{58,64}$Ni isotope beams and a Be target have been simulated using the Heavy Ion Phase Space Exploration and the Antisymmetrized Molecular Dynamics models. The present study mainly focuses on the model predictions for the excitation energies of the hot fragments and the cross sections of the final fragments produced in these reactions. The effects of various factors influencing the final fragment cross sections, such as the choice of the statistical decay code and its parameters have been explored. The predicted fragment cross sections are compared to the projectile fragmentation cross sections measured with the A1900 mass separator. At $E/A=140$ MeV, reaction dynamics can significantly modify the detection efficiencies for the fragments and make them different from the efficiencies applied to the measured data reported in the previous work. The effects of efficiency corrections on the validation of event generator codes are discussed in the context of the two models. 
\end{abstract}



\maketitle

%%%%%%%%%%%%%%%%%%%%%%%%%%%%%%%%%%%%%%%%%%%%%%%%%%%%%%%%%%%%%%%%%%%%%%%%%%%%%%%%%%%%%%%%%%%%%%%%%%%%%%%%%%%%%%%%%%%%%%%%%%
\section{Introduction}
%%%%%%%%%%%%%%%%%%%%%%%%%%%%%%%%%%%%%%%%%%%%%%%%%%%%%%%%%%%%%%%%%%%%%%%%%%%%%%%%%%%%%%%%%%%%%%%%%%%%%%%%%%%%%%%%%%%%%%%%%%%
Projectile fragmentation is a well-established technique to produce beams of exotic nuclides used for various studies in fundamental nuclear physics. It is deployed in many facilities around the world \cite{gei92,kub92,she02}. Even though the first pioneering experiments were done in the late 1970s at Berkeley \cite{hec71,gru71}, and the fragmentation process is a fundamental decay mode of highly excited nuclear systems \cite{lyn87}, the fragmentation reaction mechanism is not yet fully understood.  While there are many puzzling aspects to this phenomenon, it does display some simplifying characteristics at high incident energies. For example, many experimental observables in peripheral collisions at high energies ($>200$ MeV/nucleon), such as mass, charge, and multiplicity distributions, vary little with energy and target material. This, so-called, limiting fragmentation behavior forms the basis of empirical parameterization of the EPAX code \cite{sum00}. It allows one to predict the mass and charge distributions of projectile fragmentation reactions. Since it lacks the physics details of the reaction mechanism, the predicted cross sections deviate rather significantly from the experimental data for very neutron-rich and very proton-rich nuclei \cite{not00,moc06,moc06a,not07}. Understanding the physics of the projectile fragmentation is not only important for rare-isotope beam production purposes, but also for understanding of fundamental nuclear physics processes involved in nuclear collisions. 

A campaign of four projectile fragmentation experiments was carried out at the National Superconducting Cyclotron Laboratory at Michigan State University during 2002--2005 with a goal to measure high-quality and comprehensive projectile fragmentation cross-section data at intermediate energy. The data from eight different reaction systems yielded more than 1400 (1379 fragments $+111$ pick-up) measured cross sections. 140-MeV/nucleon \cite{moc06a} primary beams of $^{40}$Ca, $^{48}$Ca, $^{58}$Ni, and $^{64}$Ni with $^{9}$Be and $^{181}$Ta targets were used in the present experimental studies. The accuracy of these measurements provides benchmark quality sets of data for testing reaction models \cite{sub07,tsa07} as well as particle transport simulation codes that are used in the design of accelerators and radiation shielding \cite{PHITS,MARS,MCNPX,ron07}. In an effort to understand the underlying physical processes in projectile fragmentation reactions, we have performed calculations using the macroscopic-microscopic Heavy Ion Phase Space Exploration (HIPSE) model \cite{lac04} and the sophisticated fully microscopic Antisymmetrized Molecular Dynamics (AMD) model \cite{ono04}. 

This paper is structured as follows. First we introduce the main features of the two reaction models used in the present study in Section \ref{section:models}.  Then the properties of hot fragments and comparisons of data and the calculated final fragments after decay are presented in Section \ref{section:results}. The effect of sequential decays and the model dependence of the decay codes are discussed in the same section.  In addition, we discuss the effects of the detector efficiency corrections when results from dynamical models are compared to data. Our results are summarized in Section \ref{section:summary}.

%%%%%%%%%%%%%%%%%%%%%%%%%%%%%%%%%%%%%%%%%%%%%%%%%%%%%%%%%%%%%%%%%%%%%%%%%%%%%%%%%%%%%%%%%%%%%%%%%%%%%%%%%%%%%%%%%%%%%%%%%%%%
\section{Reaction Models}\label{section:models}
Understanding of the reaction dynamics of fragment production in projectile fragmentation requires reaction models more sophisticated than phenomenological ones such as EPAX \cite{sum00}, or the widely used Abrasion-Ablation (AA) model \cite{bow73,gai91}. In the first step of the AA model, collisions of spherical projectile and target nuclei are assumed. Nucleons in the overlap region are ``abraded'' and their number depends on the impact parameter. In the second step, the excited primary fragments decay. The model does not provide a mechanism to calculate the excitation energy. For simplicity, the excitation energy is assumed to be proportional to the number of abraded nucleons and sometimes adjusted to reproduce the experimental data. The model's predictive power is also limited to cross sections and cannot describe the velocity or momentum of the produced fragments. To understand the dynamics of the reactions, we employ the Heavy Ion Phase Space Exploration and Antisymmetrized Molecular Dynamics models, both of which describe the dynamical evolution of the fragments during the collision. In the present work, there is no attempt to fit the data by varying the model parameters, the nominal recommended values of the input parameters are used in both calculations.
%%%%%%%%%%%%%%%%%%%%%%%%%%

\subsection{Heavy Ion Phase Space Exploration model}
The Heavy Ion Phase Space Exploration (HIPSE) model has been implemented to bridge the gap between the statistical models, which reduce the description of the reaction to a few important parameters, and fully microscopic models \cite{lac04,lac06}. Based on a macroscopic-microscopic ``phenomenology,'' it accounts for both dynamical and statistical aspects of nuclear collisions. The HIPSE model has been shown to describe central and semi-peripheral collisions well. On the other hand, very peripheral reaction mechanisms such as knock-out, break-up, or pick-up reactions, which require the inclusion of the intrinsic quantum nature of nucleons, cannot be accounted for.

Nuclear reaction, as described by the HIPSE model \cite{lac04}, can be separated into three stages: approach of the projectile and the target nuclei, partition (formation of fragments), and the cluster propagation phase (with an in-flight statistical decay). Classical two-body dynamics of the center of masses of the target and the projectile nuclei is assumed in the entrance channel. The macroscopic proximity potential, giving a realistic Coulomb barrier, is used to describe the nucleus-nucleus potential at large distances. At small relative distances, the nucleus-nucleus potential should become sharper when the beam energy increases. To account for this effect, a phenomenological parameter, denoted by $\alpha_a$, has been introduced which extrapolates from the adiabatic limit ($\alpha_a\leq 0$) to the sudden approximation ($\alpha_a=1$). At the minimal distance of approach, nucleons in each nucleus are sampled according to a realistic zero temperature Thomas-Fermi distribution. The participant and spectator regions are then obtained using simple geometrical considerations. Nucleons outside the overlap region define the Quasi-Projectile and Quasi-Target spectators. Then two physical effects, namely direct nucleon-nucleon collisions and nucleon exchange, are treated in a simple way. When the beam energy increases, the effect of direct nucleon-nucleon collisions becomes increasingly important. This effect is modeled by assuming that a fraction, $x_{coll}$, of the nucleons in the overlap region undergoes in-medium collisions. The main effect of the in-medium collision is to slightly distort the Fermi motion hypothesis in the sampling. Once the direct collisions are over, a fraction, $x_{ex}$, of the nucleons in the overlap region is exchanged between the two spectator nuclei, relaxing the pure participant-spectator picture. After these preliminary steps, clusters are formed using a coalescence algorithm \cite{lac04} and are propagated according to a classical Hamiltonian using the same nucleus-nucleus potential as in the approach phase. To incorporate the physics of low energy reaction (below the Fermi energy), after a time denoted by $t_{froz}$, two fragments with relative separation less than their fusion barrier distance fuse if their relative energy is below their Coulomb barrier. This feature leads, in general, to a large Final State Interaction (FSI). Once all the FSIs are processed, the nuclei cannot exchange particles anymore and a chemical freeze-out is reached. At this stage, the total excitation energy can be determined event-by-event from the energy conservation and assigned to clusters, which finally undergo in-flight decay. 

The HIPSE model has only three adjustable parameters ($\alpha_a$, $x_{ex}$, $x_{coll}$). The values of these parameters have been adjusted \cite{lac06} for beam energies of 10, 25, 50, 80 \mevu. Using simple functions, we extrapolated the values of $\alpha_a=0.55$, $x_{ex}=0.09$, and $x_{coll}=0.18$ \cite{moc06} to our beam energy of 140 \mevu. In order to compare the HIPSE simulation with results from the Antisymmetrized Molecular Dynamics (AMD) model (see below), the time, $t_{froz}$, originally taken as 50 fm/c has been increased to 150 fm/c. We have checked that this does not affect the final results.

The HIPSE model originally includes the in-flight decay based on an improved version of the SIMON decay model \cite{dur92}. In the present study, the phase space generated by HIPSE before the in-flight decay is used as input to the GEMINI decay code \cite{cha88} which is known to give a better treatment of sequential decays of excited nuclei. As a consequence, some spatial-temporal correlations may be lost. The influence of the decay code will be discussed in more detail below.

%%%%%%%%%%%%%%%%%%%%%%%%%%%%%%%%%%%
\subsection{Antisymmetrized Molecular Dynamics model}
The Antisymmetrized Molecular Dynamics (AMD) model \cite{ono04,ono92} has been chosen from among many microscopic models to simulate the fragmentation reactions measured in our experiments. As one of the most sophisticated transport models, it describes the nuclear reaction at the microscopic level of interactions of individual nucleons. In the AMD model, a potential is used to take into account all reaction processes involved in the complex heavy-ion collisions.

The AMD wave function is given by a Slater determinant of Gaussian wave packets for individual nucleons. Centroids of these wave packets are treated as dynamical variables.  An effective nuclear interaction determines the one-body motion of the wave packets by the mean field. The correlations are introduced by expressing the many-body state as an ensemble of many AMD wave functions, i.e., by adding stochastic terms to the equation of motion. Nucleon-nucleon scattering is included as a stochastic process.  The probabilities of collisions are determined by the assumed in-medium cross sections of nucleon-nucleon collisions. Furthermore, another stochastic term is considered in order to take into account the change of the width and shape of the phase-space distributions of individual nucleons. The single-particle wave functions in each channel are Gaussian wave packets with a fixed width parameter, which is advantageous in describing the fragment formation.  It should be noted that the time evolution is solved independently for each channel, and the interference of different channels is neglected.

In the present study, we employ a Gogny-type force (Gogny-AS \cite{ono03}) as the effective nucleon-nucleon interaction and the free two-nucleon collision cross sections with a cut-off at 150 mb as the in-medium nucleon-nucleon cross section.

The ground states of $^{40,48}$Ca and $^{58,64}$Ni projectiles and $^{9}$Be target were prepared by the frictional cooling method \cite{ono92} applied to the AMD wave function. The AMD simulations were carried out for an impact parameter range of 0--10 fm and up to the time of 150 fm/c, when the primary fragments are spatially well separated. These fragments, recognized by a simple coalescence algorithm, with the associated excitation energy \cite{moc06} are decayed using the GEMINI code \cite{cha88}.

%%%%%%%%%%%%%%%%%%%%%%%%%%%%%%%%%%%%%%%%%%%%%%%%%%%%%%%%%%%%%%%%%%%%%%%%%%%%%%%%%%%%%%%%%%%%%%%%%%%%%%%%%%%%%%%%%%%%%%%%%%%%%%%%
\section{Results}\label{section:results}
%%%%%%%%%%%%%%%%%%%%%%%%%%%%%%%%%%%%%%%%%%%%%%%%%%%%%%%%%%%%%%%%%%%%%%%%%%%%%%%%%%%%%%%%%%%%%%%%%%%%%%%%%%%%%%%%%%%%%%%%%%%%%%%%%%
\subsection{Primary fragments}
\begin{figure}
\begin{center}
\includegraphics[width=.9\textwidth]{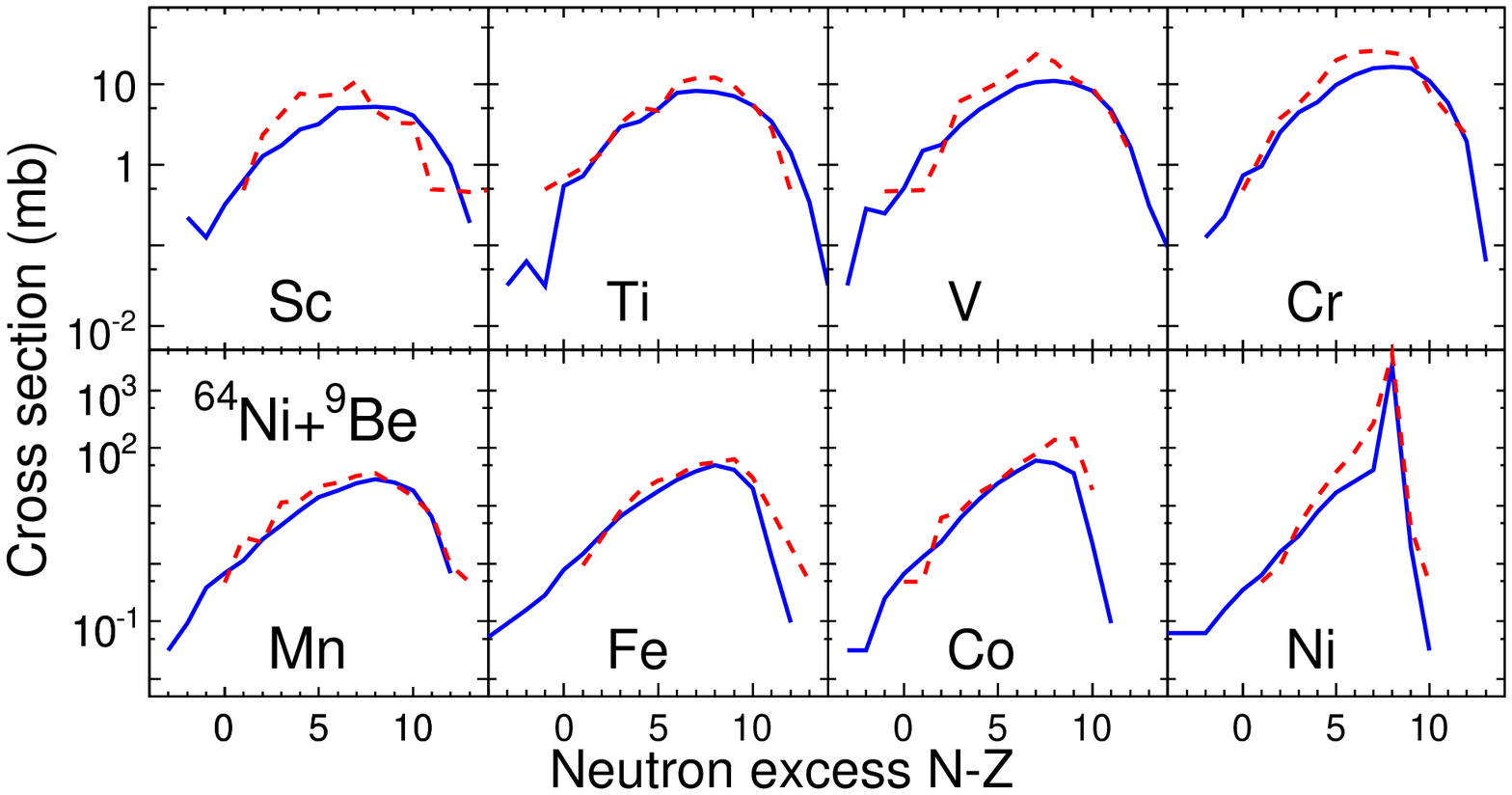}
\caption{(Color online) Primary fragment isotopic distributions for the $^{64}$Ni+$^9$Be  reaction system plotted as a function of neutron excess, $N-Z$. Solid and dashed lines show calculations by HIPSE and AMD models, respectively.}\label{fig1}
\end{center}
\end{figure}

The most direct comparison of different model calculations is to examine the properties of the primary fragments before sequential decays occur. Fig. \ref{fig1} presents the isotopic primary fragment distributions obtained by the HIPSE (solid line) and AMD (dashed lines) models for the $^{64}$Ni+$^9$Be reaction system. In general, the total fragment cross sections from AMD are higher resulting in higher isotope cross sections around the peak compared to HIPSE results. For elements close to the projectile ($Z\ge 25$), both models predict very similar isotope distributions with relatively narrow widths (bottom panels). The models start to show increasing differences in the widths, the centroids, and the magnitudes of the cross sections for the isotope distributions with increasing number of removed protons. With more removed nucleons (upper panels), the centroids of the isotopic distributions from AMD are shifted to less neutron-rich isotopes. As expected, the shifts are more pronounced for neutron-rich projectiles of $^{48}$Ca and $^{64}$Ni \cite{moc06}. In the AMD model, the centroids and widths of the isotope distributions are expected to depend on the symmetry energy terms of the effective interaction, as is the case for central collisions \cite{ono04b,ono07}. In the HIPSE model, experimental masses and empirical formula are involved in the computation of Q values and excitation energy but no explicit density dependence of the symmetry energy is included.

%%%%%%%%%%%%%%%%%%%%%%%%%%%%%%%%%%%%%%%%%%%%%
\subsection{Excitation energy}
After nuclear collisions, the excited projectile-residue decays through emission of light particles. The evolution of the decaying system depends on the excitation energy transferred. However, in projectile fragmentation experiments we detected fragments in forward angles ($\pm 30$ mrad) with velocities close to that of the projectile \cite{moc06}. The mean excitation energy can be deduced only indirectly from theoretical models. In the simple Abrasion-Ablation (AA) model \cite{bow73}, the excitation energy of primary fragments is assumed to be proportional to the number of nucleons removed \cite{fri00,gai91,moc06}. To best describe fragmentation cross sections, both the excitation energy and its fluctuations are determined by fitting the data \cite{moc06}. Contrary to the AA models, the models considered here calculate the excitation energy and its fluctuations. The HIPSE model defines the excitation energy of the residues from energy conservation, and the AMD calculation defines the excitation energy by evaluating the expectation value of the Hamiltonian for the many-body wave function of the primary fragment. 

\begin{figure}
\begin{center}
\includegraphics[width=.4\textwidth]{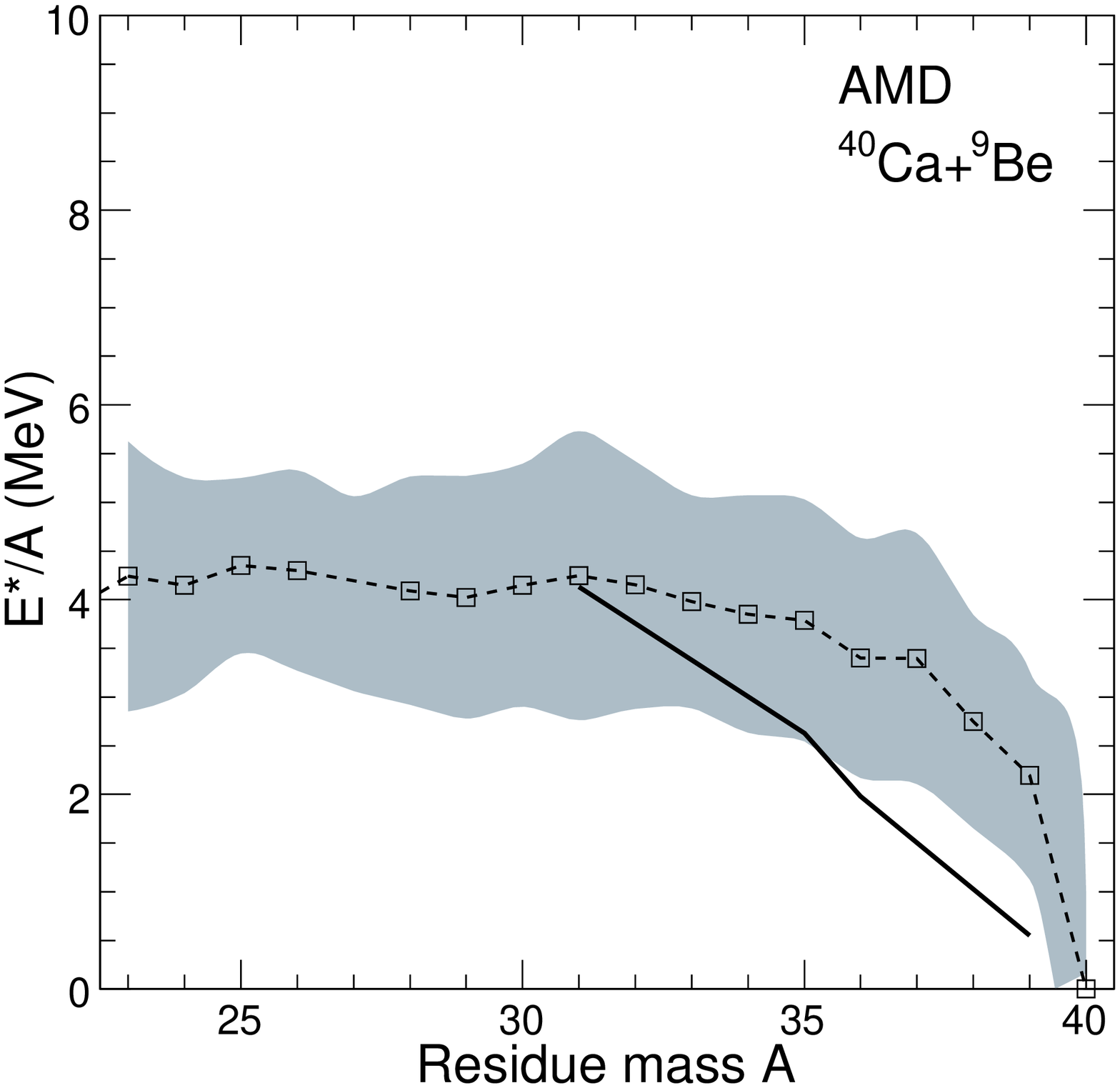}
\includegraphics[width=.4\textwidth]{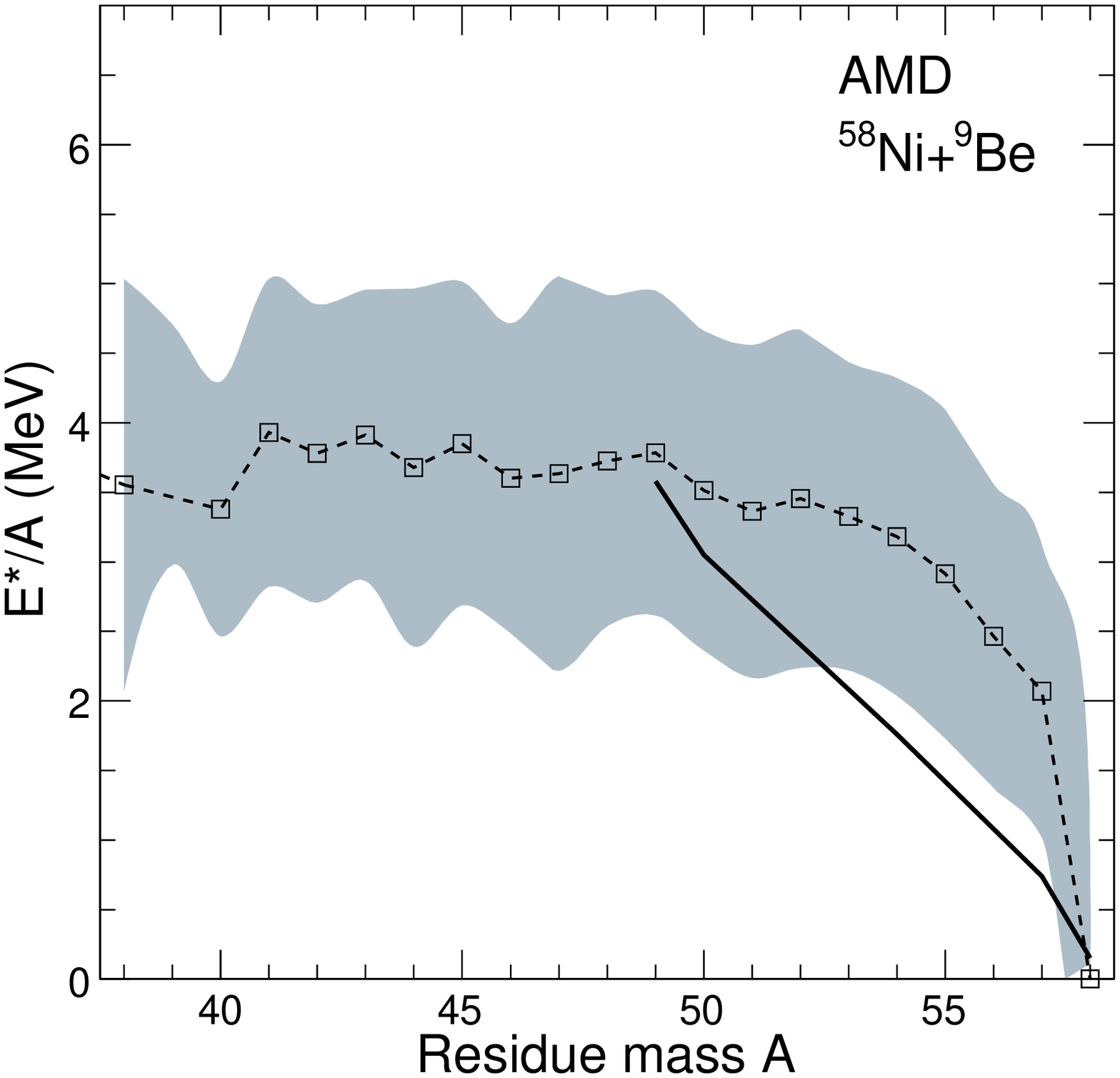}
\includegraphics[width=.4\textwidth]{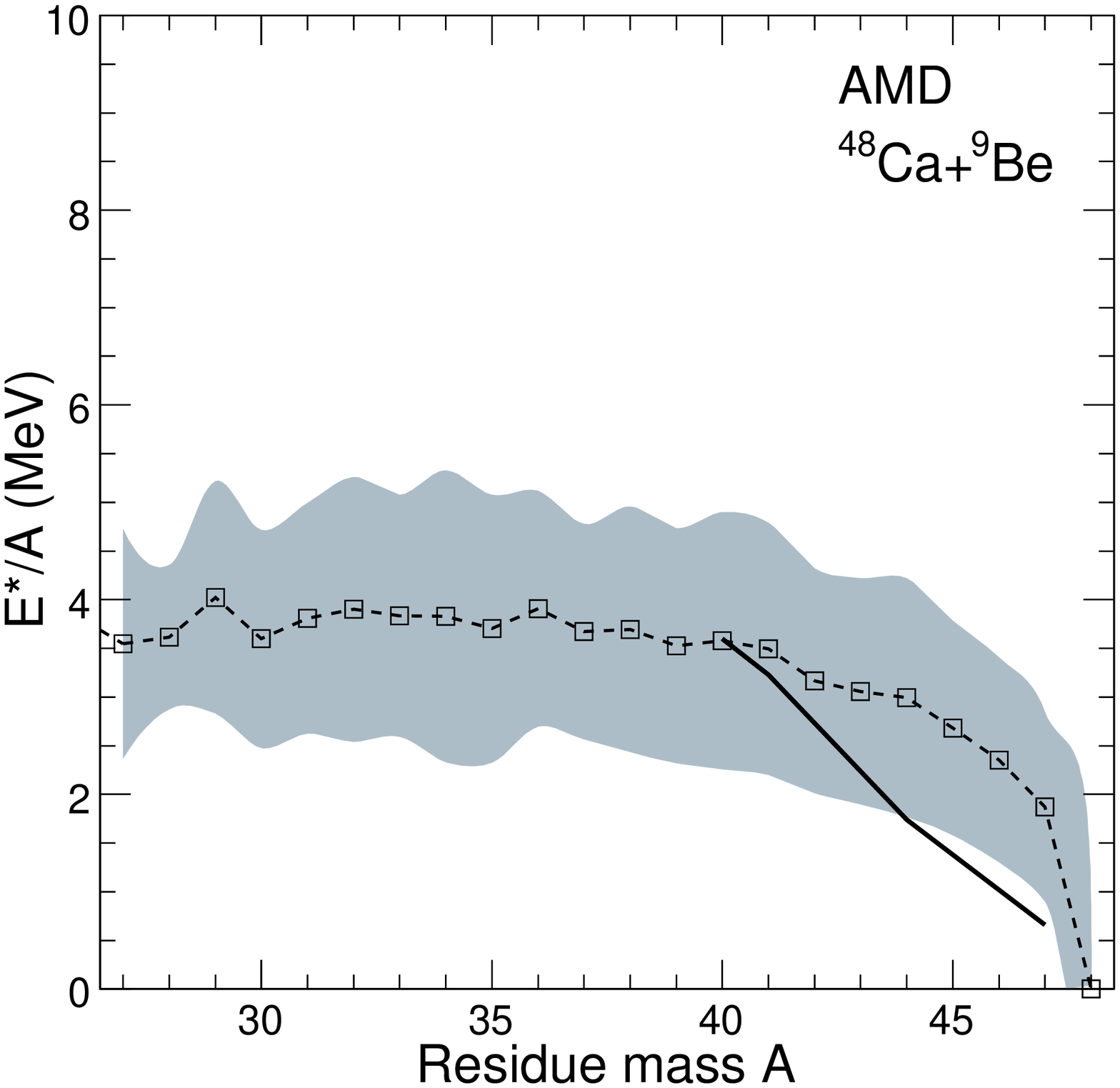}
\includegraphics[width=.4\textwidth]{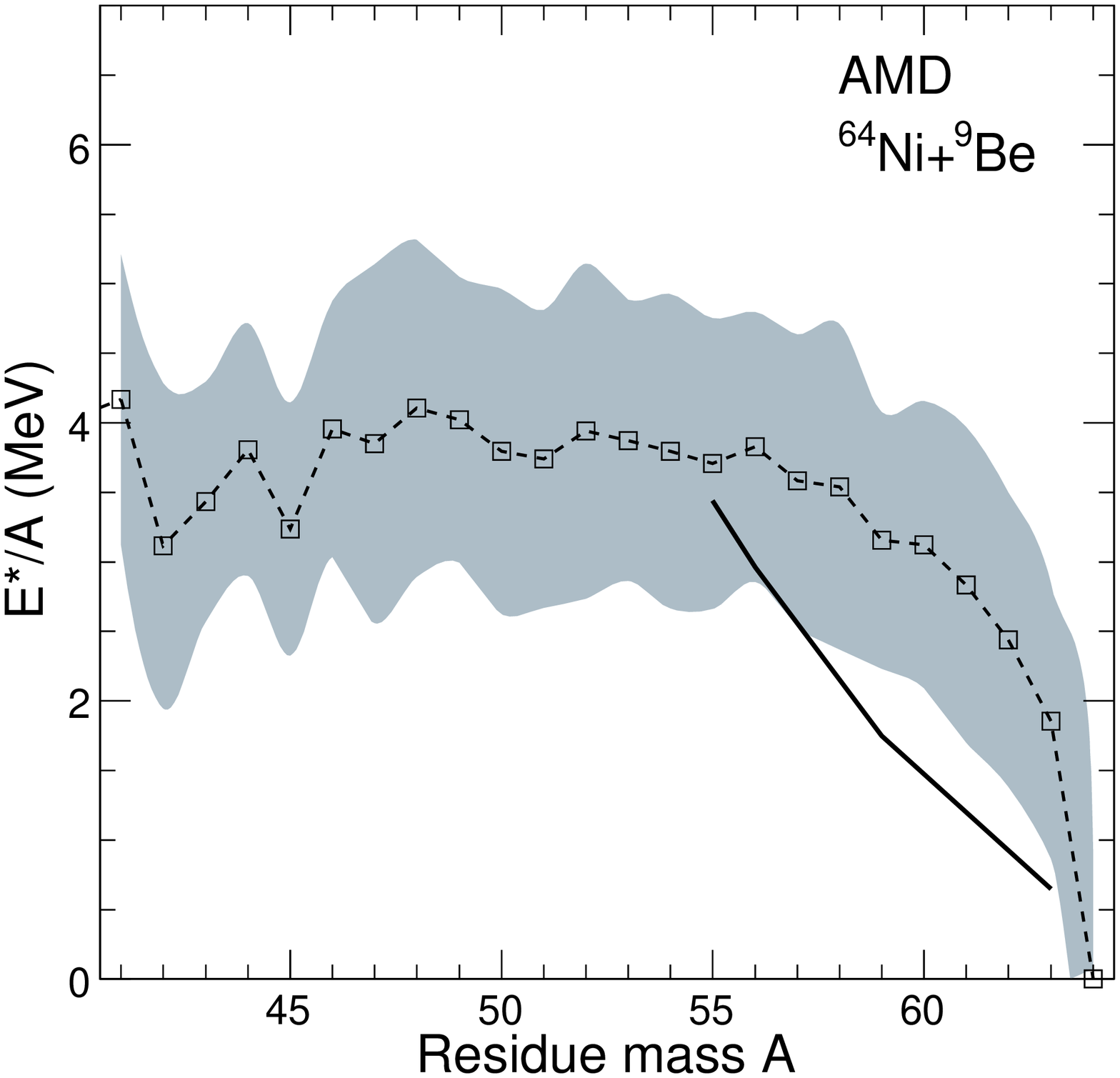}
\caption{(Color online) The mean excitation energy per nucleon, $E^\ast/A$, plotted as a function of the mass number of the primary fragments for reactions of $^{40}$Ca (top left), $^{48}$Ca (bottom left), $^{58}$Ni (top right), and $^{64}$Ni (bottom right) with $^9$Be target. The mean excitation energy per nucleon calculated by the AMD model is shown as open squares connected by the dashed line to guide the eye. The shaded region depicts fluctuations in the excitation energy per nucleon expressed in terms of one standard deviation around the mean. For reference, the BUU calculation results are plotted as a solid line.}\label{fig2}
\end{center}
\end{figure}

\begin{figure}
\begin{center}
\includegraphics[width=.4\textwidth]{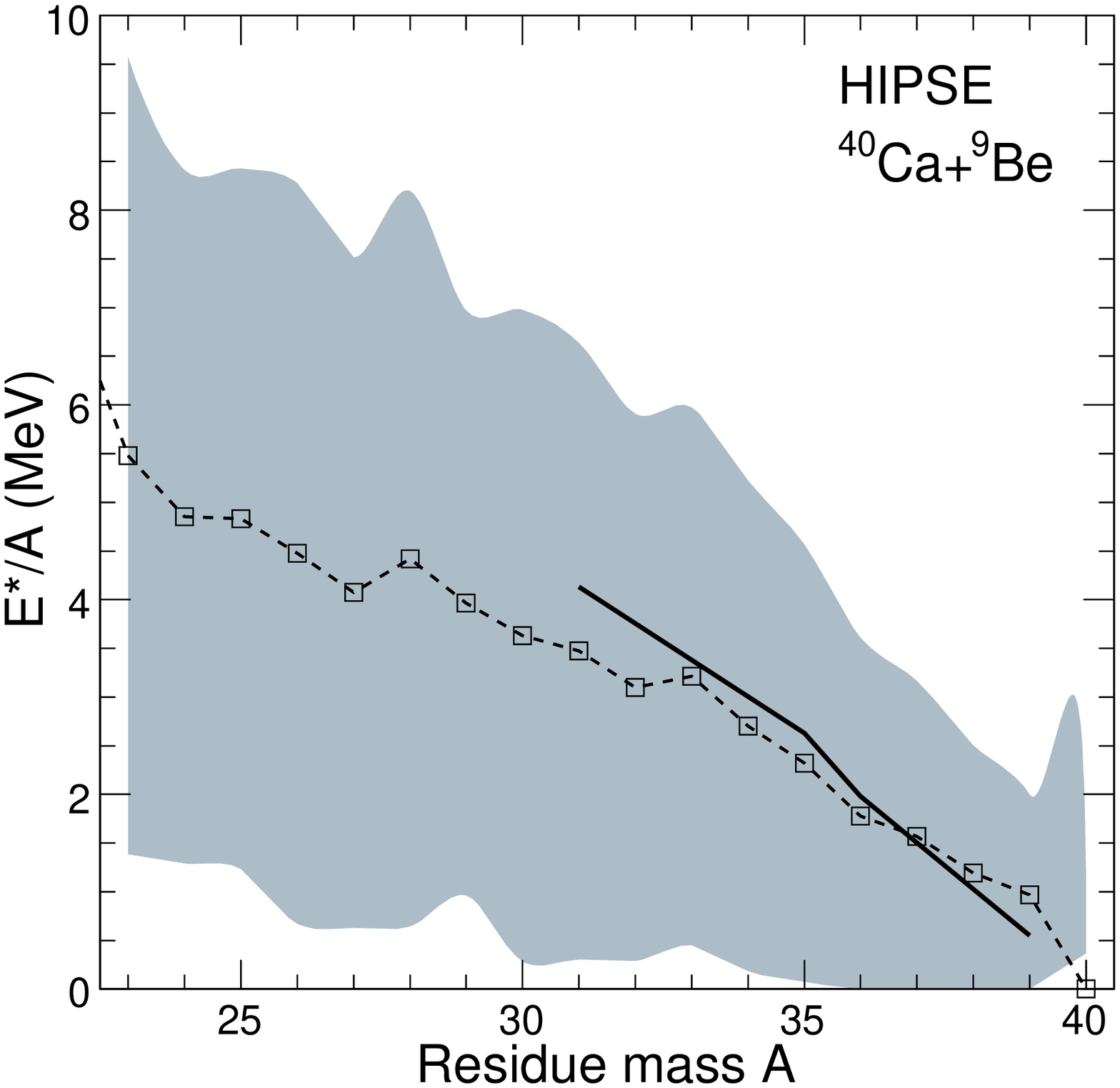}
\includegraphics[width=.4\textwidth]{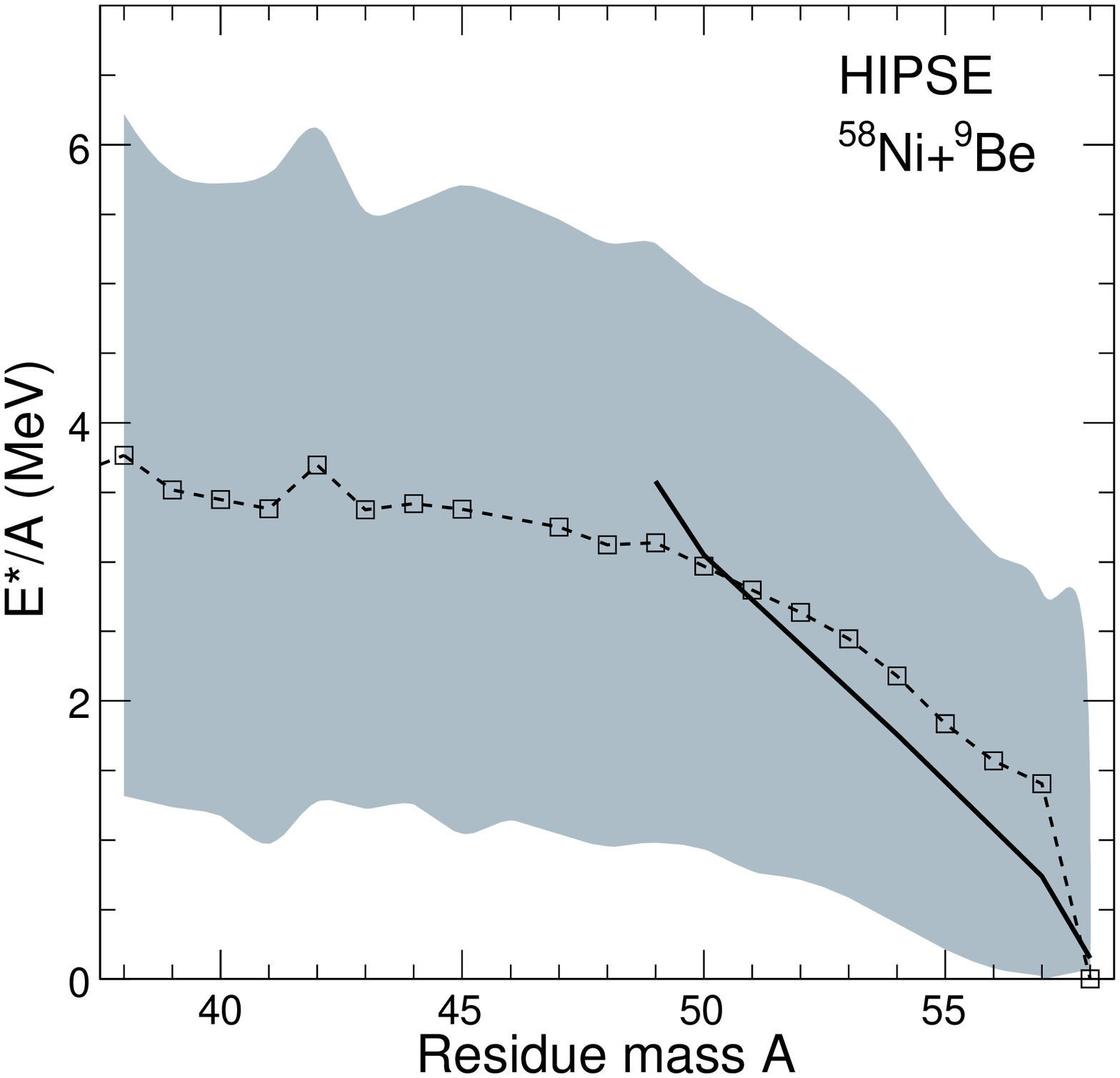}
\includegraphics[width=.4\textwidth]{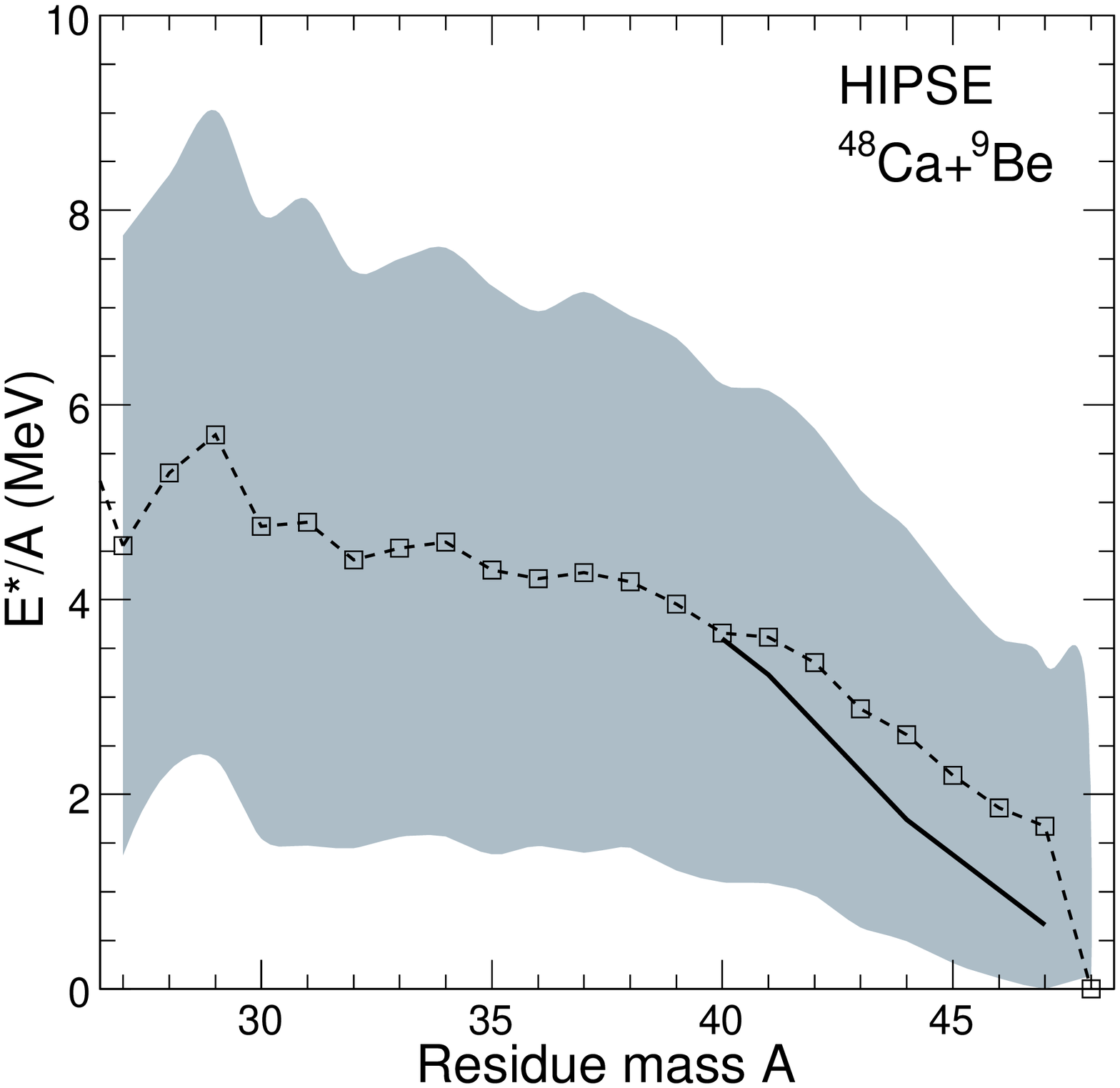}
\includegraphics[width=.4\textwidth]{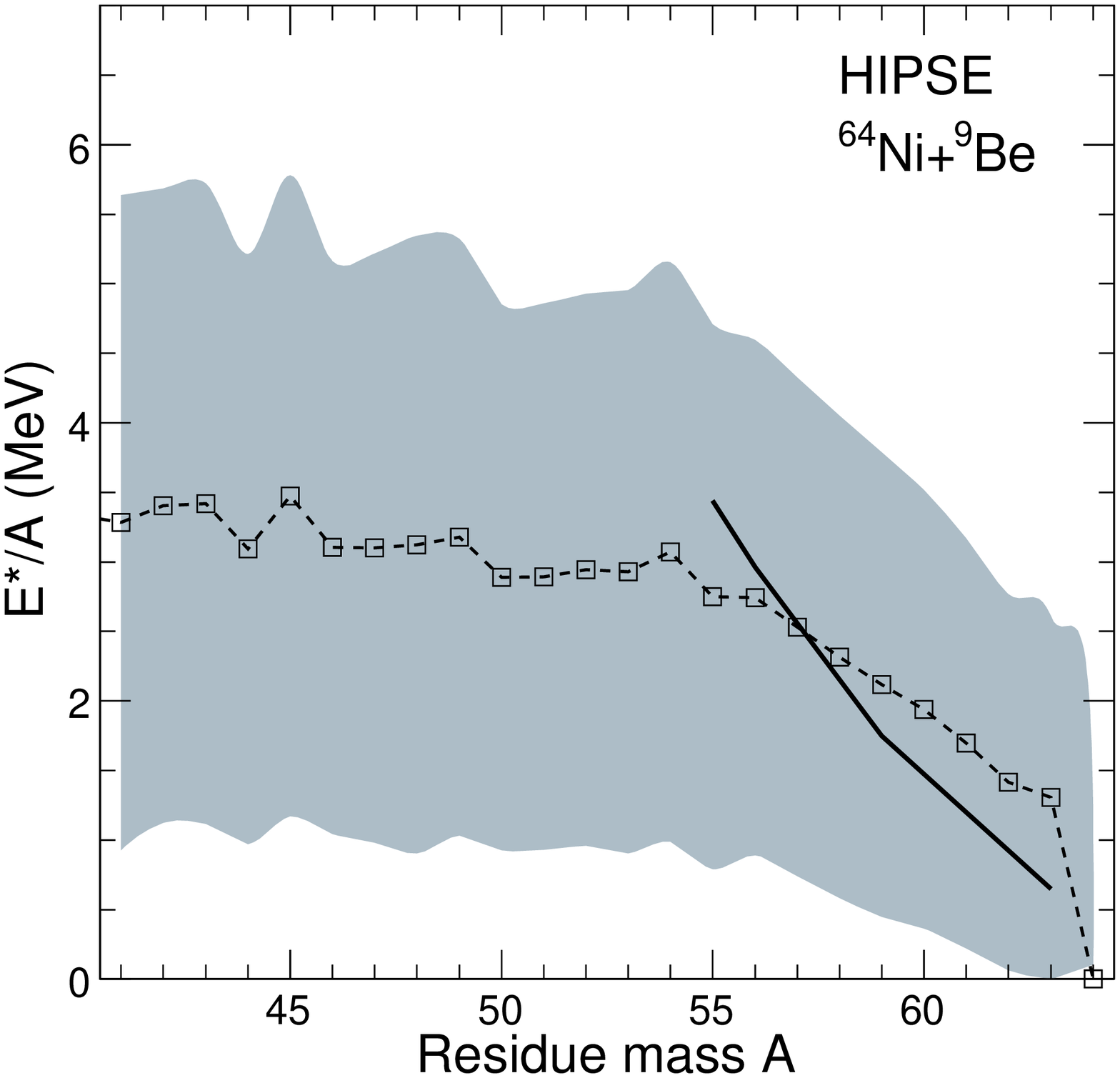}
\caption{(Color online) The mean excitation energy per nucleon, $E^\ast/A$, plotted as a function of the mass number of the primary fragments. Same convention as Fig.~\ref{fig2} is used.}\label{fig3}
\end{center}
\end{figure}

The average excitation energy per nucleon, $E^\ast/A$, for primary fragments produced in the fragmentation of $^{40,48}$Ca and $^{58,64}$Ni on $^{9}$Be target is shown as open squares in Fig. \ref{fig2} and \ref{fig3} for the AMD and HIPSE calculations, respectively, as a function of the primary fragment mass number. The shaded regions show the root-mean-square (RMS) widths of the excitation energy distributions. For reference, we calculated the residue excitation energy with a single particle, microscopic Boltzman-Uehling-Uhlenbeck (BUU) equation \cite{dan00}. The BUU results are deterministic and give average values of the observable for a given impact parameter. They are shown as solid lines in Fig.~\ref{fig2} and \ref{fig3}. The BUU results which do not extend to small residue masses, exhibit trends more similar to the HIPSE calculations.

In the case of the AMD simulations, we notice a rather sharp rise of $E^\ast/A$ with the number of removed nucleons close to the projectile. The excitation energy saturates around 4 MeV after removal of about 10 nucleons. The saturation is inconsistent with the assumptions used in AA models, which assume that excitation energy is proportional to the number of abraded nucleons. In the HIPSE model, similar saturation values are obtained especially in the case of the Ni isotopes. However, the excitation energy fluctuations are much larger in the HIPSE model than the AMD model.

For residues close to the projectile ($\approx 0$--10 abraded nucleons), the AMD calculation produces systematically higher  excitation energy. This could be due to different cluster formation in the models. The simple nucleon sampling procedure used in HIPSE to construct fragments may predict lower excitation for projectile-like fragments. As discussed in Section~\ref{section:fragmentCS} the excitation energy profile may be related to the widths of the isotope distributions.

Saturation of the excitation energy is not obvious in the fragmentation of the Ca isotopes in the HIPSE calculations. The residue excitation energy increases with decreasing masses but the rate of increase is much less for residues lighter than 30 for $^{40}$Ca and 38 for $^{48}$Ca projectiles. Monotonic increase is observed in the reactions with $^{181}$Ta targets in the HIPSE calculations. The latter increase may be related to the increase of the maximum number of nucleons involved in the collisions and exchange with the quasi-projectile when the targets are changed from $^9$Be to $^{181}$Ta. However, different profiles of the mean excitation energy for the $^9$Be and $^{181}$Ta are not supported by the data, which show very little target dependence \cite{moc06}.  Unfortunately, we could not perform the AMD calculations involving $^{181}$Ta targets to check the observations seen in the HIPSE model calculations because of excessive CPU time required.

%%%%%%%%%%%%%%%%%%%%%%%%%%%%%%%%%%%%%%%%%%%%%%%%
\subsection{Evaporation Codes}
The primary fragments produced by different reaction models cannot be directly compared to the experimental data, because the experimental observation of fragments is performed after hundreds of nanoseconds, many orders of magnitude later, than the prompt step simulated by the nuclear reaction models (HIPSE, AMD). Direct comparison with data requires incorporating the sequential decay models as the second step after excited fragments are formed. 

Currently, there are no standardized sequential decay codes \cite{tsa06}. The GEMINI code, widely used in performing sequential decay of hot fragments, calculates the decay of a primary fragment by sequential binary decays. Monte Carlo technique is employed to follow all decay chains until the resulting products are unable to undergo further decay. For the purposes of the sequential decay calculations the excited primary fragments generated by the HIPSE and AMD model calculations are taken as the compound nucleus \cite{boh36} input to the GEMINI code. Hence, every primary fragment is decayed as a separate event \cite{moc06}. For the evaporation of particles lighter than an alpha-particle, the Hauser-Feshbach \cite{hau52} formalism is applied. The liquid drop model with shell corrections \cite{kra79} is used to calculate the masses of all parent and daughter nuclei in the calculation. The Fermi gas \cite{bet36} expression is used to calculate the level density.

\begin{figure}
\begin{center}
\includegraphics[width=.8\textwidth]{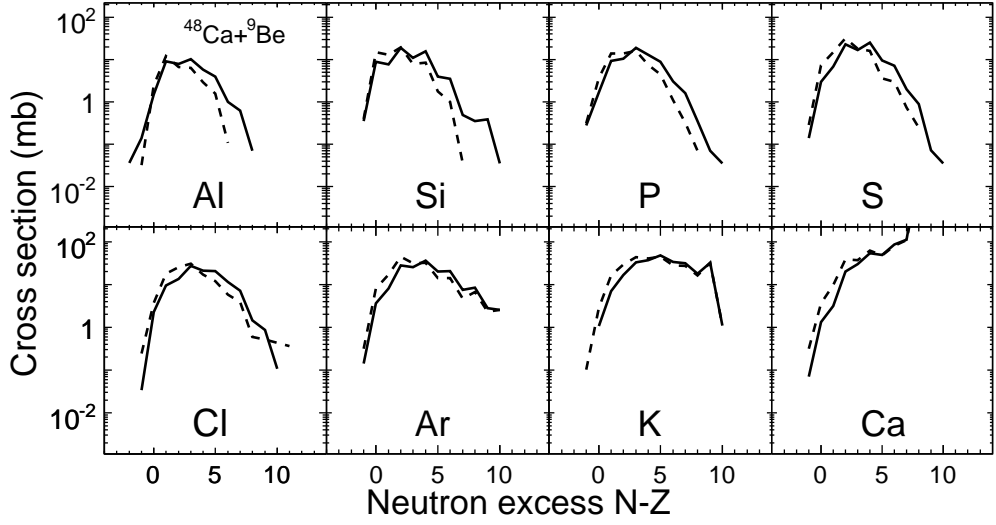}
\caption{Isotopic cross-section distributions for $^{48}$Ca+$^9$Be for $13\leq Z\leq 20$ elements calculated using the AMD model coupled to GEMINI decay calculations with level density parameter $a=A/8$ MeV$^{-1}$ (dashed line) are compared to calculations with $a=A/12$ MeV$^{-1}$ (solid line).}\label{fig4}
\end{center}
\end{figure}

For neutron-rich projectiles, the final isotope distributions are found to be sensitive to the selection of the level density parameter, $a$. To demonstrate this effect, results from the hot fragments produced in AMD simulations of the collisions of the $^{48}$Ca beam on a $^{9}$Be target using two different level density parameterizations are shown in Fig. \ref{fig4}. The solid lines are calculated using $a = A/12$ MeV$^{-1}$ and the dashed lines depict the calculation using $a = A/8$ MeV$^{-1}$. The calculation using $a=A/12$ MeV$^{-1}$ results in wider isotope distributions and shifting of the peaks towards the more neutron-rich isotopes. Such shifts are also observed in the case of  the neutron-rich $^{64}$Ni beam \cite{moc06}. The change of the level density parameter while keeping all other parameters of the sequential decay constant corresponds to an effective change of temperature of the decaying system. In this picture, decaying compound nucleus with higher temperature ($a = A/12$ MeV$^{-1}$) leads to wider isotope distributions with more neutron-rich fragments. On the other hand, the decay of the less neutron-rich fragments such as those produced in the $^{40}$Ca and $^{58}$Ni induced reactions will not be affected very much.  Unless otherwise noted, all sequential decay calculations in this paper use $a = A/10$ MeV$^{-1}$.

\begin{figure}
\begin{center}
\includegraphics[width=.8\textwidth]{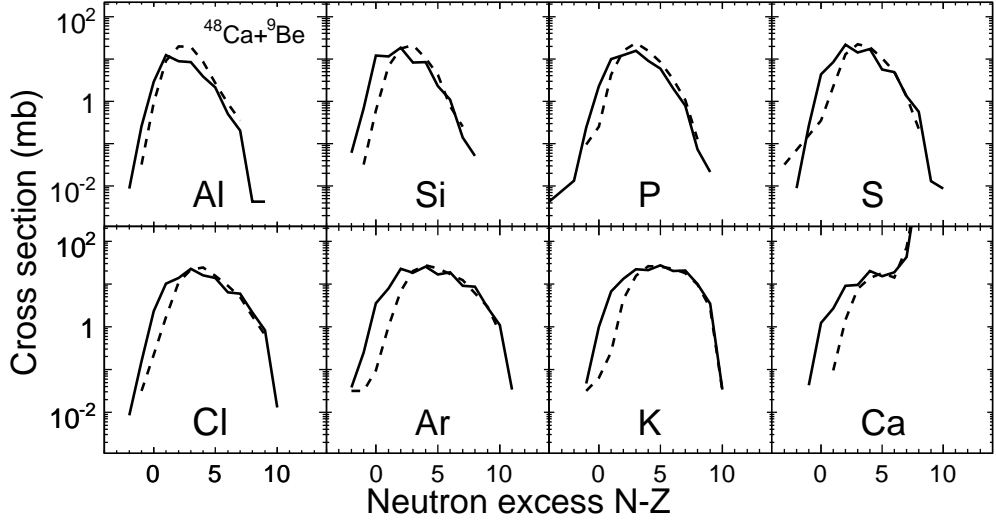}
\caption{Isotopic cross-section distributions of $^{48}$Ca+$^9$Be reactions calculated by HIPSE with GEMINI decay  (solid line) are compared to HIPSE coupled to SIMON decay (dashed line).}\label{fig5}
\end{center}
\end{figure}

In addition to the level density parameter, the final fragment distributions also depend on the evaporation code used \cite{tsa06}. For example, if the SIMON decay code is used instead of GEMINI, the fragment distributions are different. The dashed lines in Fig. \ref{fig5} are predictions from HIPSE coupled with the decay code SIMON while the solid lines are predictions from HIPSE using GEMINI as the decay code for the $^{48}$Ca+$^9$Be reaction. In general, results from GEMINI reproduce the cross sections of the neutron deficient fragments consistently better than calculations using SIMON to decay the hot fragments. Detailed comparisons of different decay codes have been published in Ref. \cite{tsa06}. In particular, SIMON code failed to exhibit the isoscaling behavior which is reproduced by other statistical codes. Nonetheless, the uncertainties introduced by different decay codes or using different parameters in the decay codes can be as large as the differences of the results between the AMD and HIPSE models.

%%%%%%%%%%%%%%%%%%%%%%%%%%%%%%%%%%%%%%%%%%%%%%%%%%%%
\subsection{Fragment cross sections}\label{section:fragmentCS}
In this section we compare the experimentally determined reaction cross sections with the final fragment cross sections predicted by the HIPSE and AMD models. To minimize the influence of different sequential decay codes, GEMINI is used as the evaporation code to decay the excited primary fragments created in both models. 

Experimentally, only fragments with velocities that match the acceptance of the fragment separator are measured. Ideally, such experimental constraints have to be compensated so that the experimental cross-section data can be compared directly with predicted cross sections from theoretical models. This is especially true for models such as the EPAX parameterization of the fragment cross sections and the abrasion-ablation models, which do not contain dynamic information about the collisions. However, the detection efficiency coefficients used in converting the measured (raw) cross sections to total cross sections depend on the assumptions of the transmission efficiency and angular distributions of the fragments. In general, fragment transmission through the magnetic spectrometer used is obtained from ion-optics simulations and the transmission efficiency is better than 95\%. On the other hand, it is very difficult to estimate the fragment angular distributions with certainty. In the experiment, only the momentum distributions accepted within the spectrograph are measured. Transverse momentum distributions cannot be measured easily and their distributions are normally estimated using parameterizations \cite{moc06}. 

\begin{figure}
\begin{center}
\includegraphics[width=.8\textwidth]{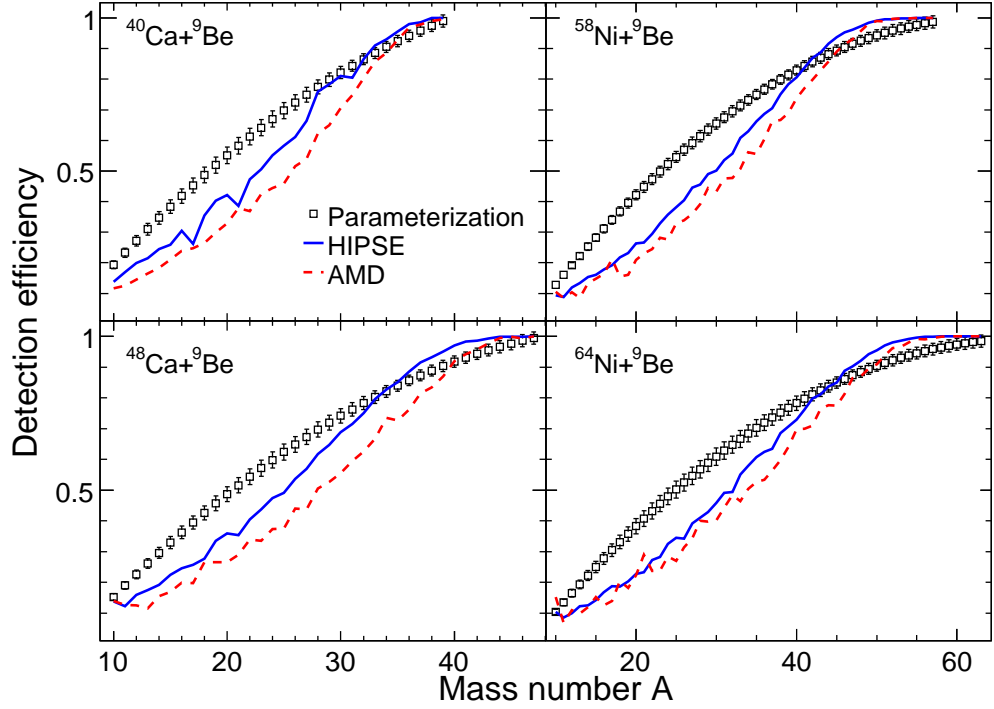}
\caption{(Color online) Transmission correction, used to correct the experimental data based on parameterization \cite{moc06a}, is plotted as a function of the fragment mass number (open squares) for all investigated reaction systems. The solid and dashed lines show the transmission correction deduced from the HIPSE and AMD calculations, respectively.}\label{fig6}
\end{center}
\end{figure}

\begin{figure}
\begin{center}
\includegraphics[width=.8\textwidth]{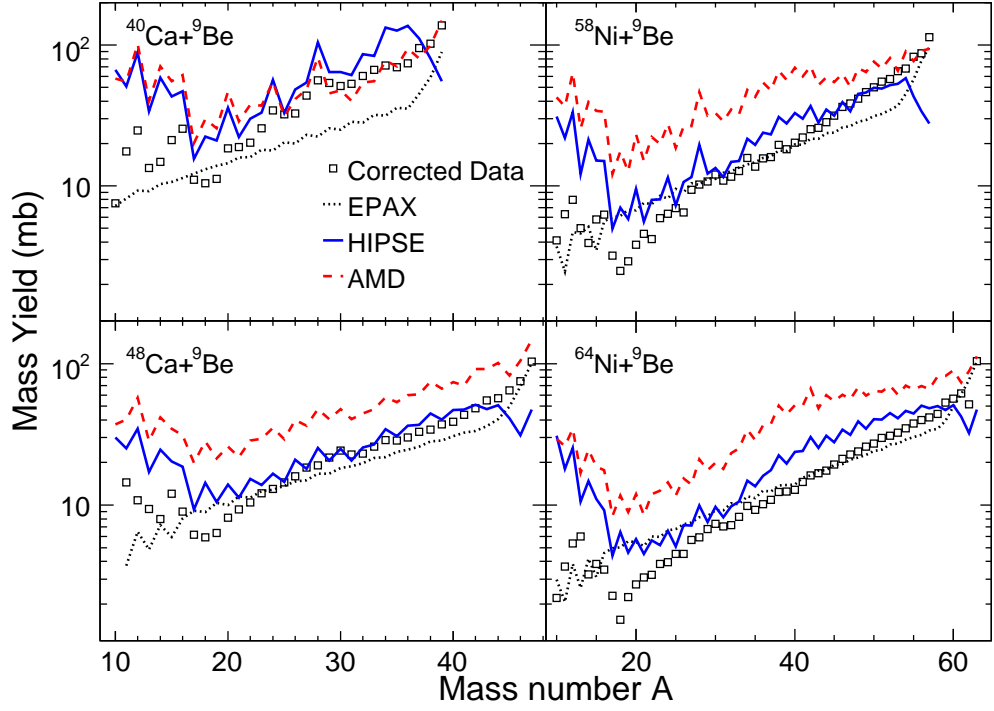}
\caption{(Color online) Mass cross-section distributions (open squares) for four reaction systems are compared to predictions from two reaction models, HIPSE (solid lines) and AMD (dashed lines).}\label{fig7}
\end{center}
\end{figure}

\begin{figure}
\begin{center}
\includegraphics[width=.8\textwidth]{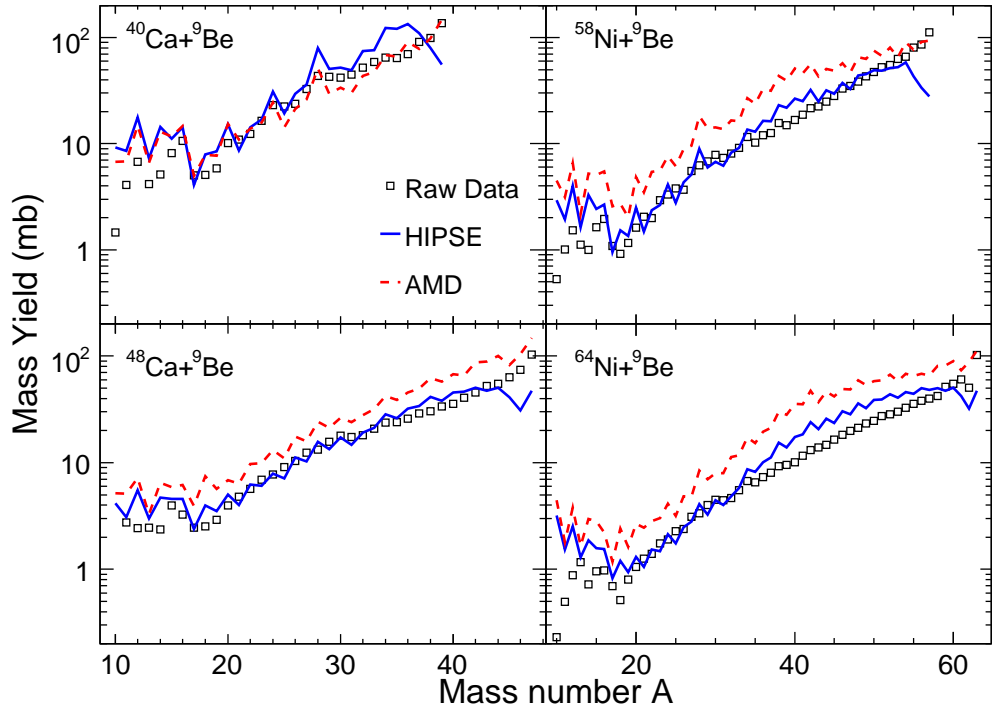}
\caption{(Color online) Uncorrected cross-section mass distributions (open squares) for four reaction systems are compared to filtered events from two reaction models, HIPSE (solid lines) and AMD (dashed lines).}\label{fig8}
\end{center}
\end{figure}

The open squares in Fig. \ref{fig6} show the mass dependence of the transmission correction factor, $\varepsilon$, used in Ref. \cite{moc06,moc06a} to obtain the published fragment cross sections for the four reactions studied here. From dynamical models, one can calculate the correction factors by constructing the ratios of fragment cross sections filtered by the experimental acceptance, $\sigma_{filtered}$, to the calculated fragment cross sections, $\sigma_{model}$: 
\begin{equation}
\varepsilon_{th}=\sigma_{filtered}/\sigma_{model}.
\end{equation}
The corresponding correction factors, $\varepsilon_{th}$, obtained from the models, a solid line for HIPSE and a dashed line for AMD, are quite different from $\varepsilon$ (open squares), used to correct the experimental data, suggesting the angular distributions assumed in $\varepsilon$ do not agree with the angular distributions described by the models. The differences are model dependent and largest for lighter fragments. Fig. \ref{fig6} illustrates that it may not be appropriate to compare calculated results directly to the published data as is customarily done \cite{PHITS,MARS,MCNPX,ron07}. As the correction factors are model dependent, it is more accurate to compare filtered calculations to uncorrected (raw) experimental cross sections. To illustrate the differences in comparing corrected data with unfiltered theoretical results (Fig. \ref{fig7}) and raw data with filtered calculations (Fig. \ref{fig8}), we plot the mass distributions for the four systems studied here. The data are shown as open symbols and lines are predictions from various models. Since the detection efficiency decreases with the mass of the detected fragments, lighter masses that are less than half of the projectile masses have the largest corrections. This is expected as the lighter masses have a larger velocity or momentum spread in nearly all models. For reference, the dotted lines in Fig. \ref{fig7} are EPAX predictions. Agreement with the EPAX results is better for the $^{48}$Ca+$^{9}$Be and $^{58}$Ni+$^{9}$Be reactions as the mass distributions of both of these reactions were used to extract the EPAX parameters \cite{sum00}. 

In Fig. \ref{fig8}, the filtered results, obtained by applying the experimental acceptance cut of 30 mrad to the simulated events from both the HIPSE and AMD models agree with the raw data much better, especially for light fragments. For the HIPSE model, the drop of fragment yields around the projectiles is due to inadequate fluctuations in the most peripheral collisions. Except for $^{40}$Ca induced reactions, the AMD model predicts higher fragment yields for the other three reactions and the HIPSE model reproduces the overall magnitude of the cross sections. 

\begin{figure}
\begin{center}
\includegraphics[width=.8\textwidth]{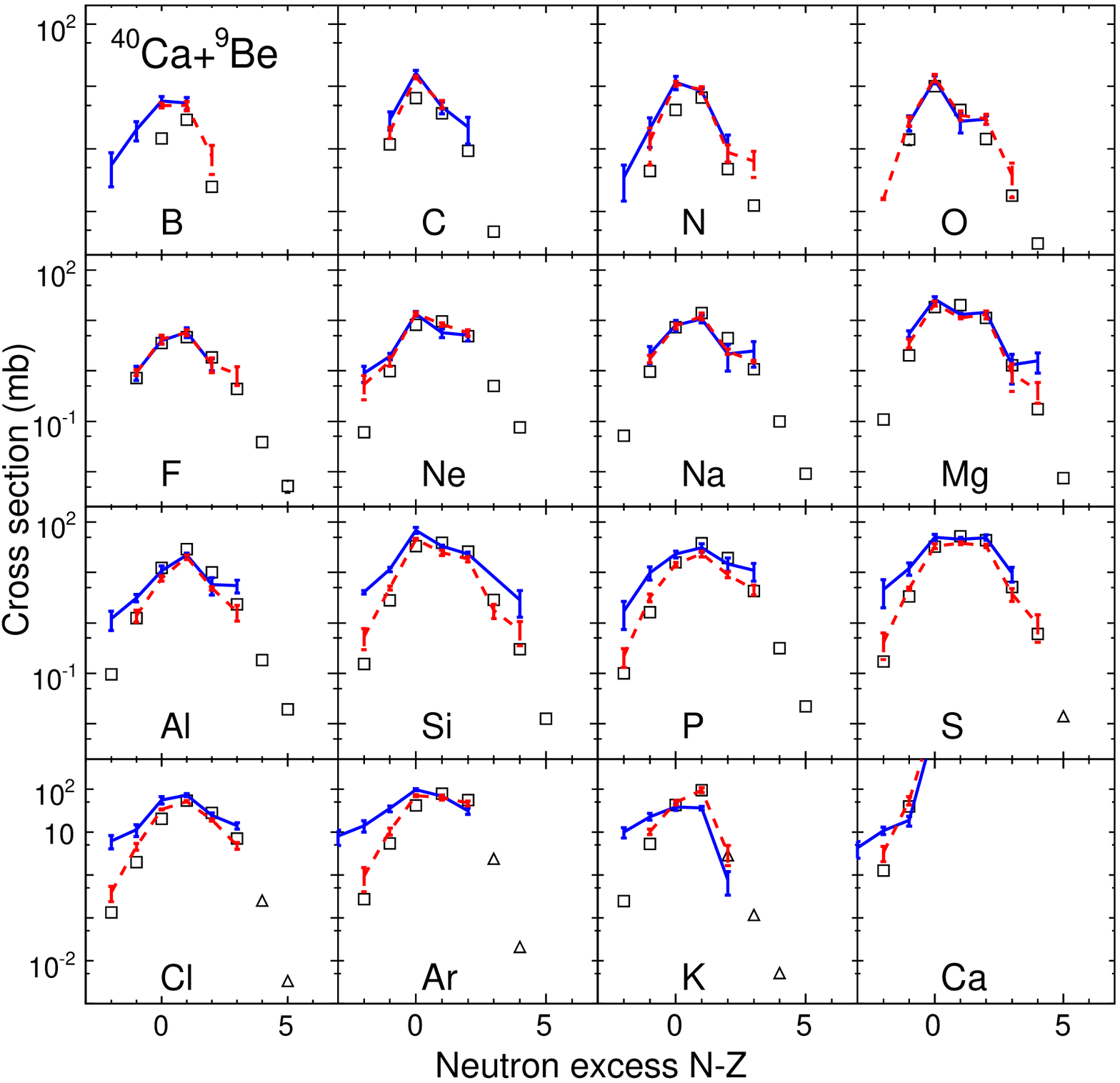}
\caption{(Color online) Fragmentation (open squares) and nucleon pick-up (open triangles) cross sections of $^{40}$Ca+$^9$Be reactions are compared to calculations by HIPSE (solid line) and AMD (dashed line) models. The error bars in the data are smaller than the symbols.}\label{fig9}
\end{center}
\end{figure}

\begin{figure}
\begin{center}
\includegraphics[width=.8\textwidth]{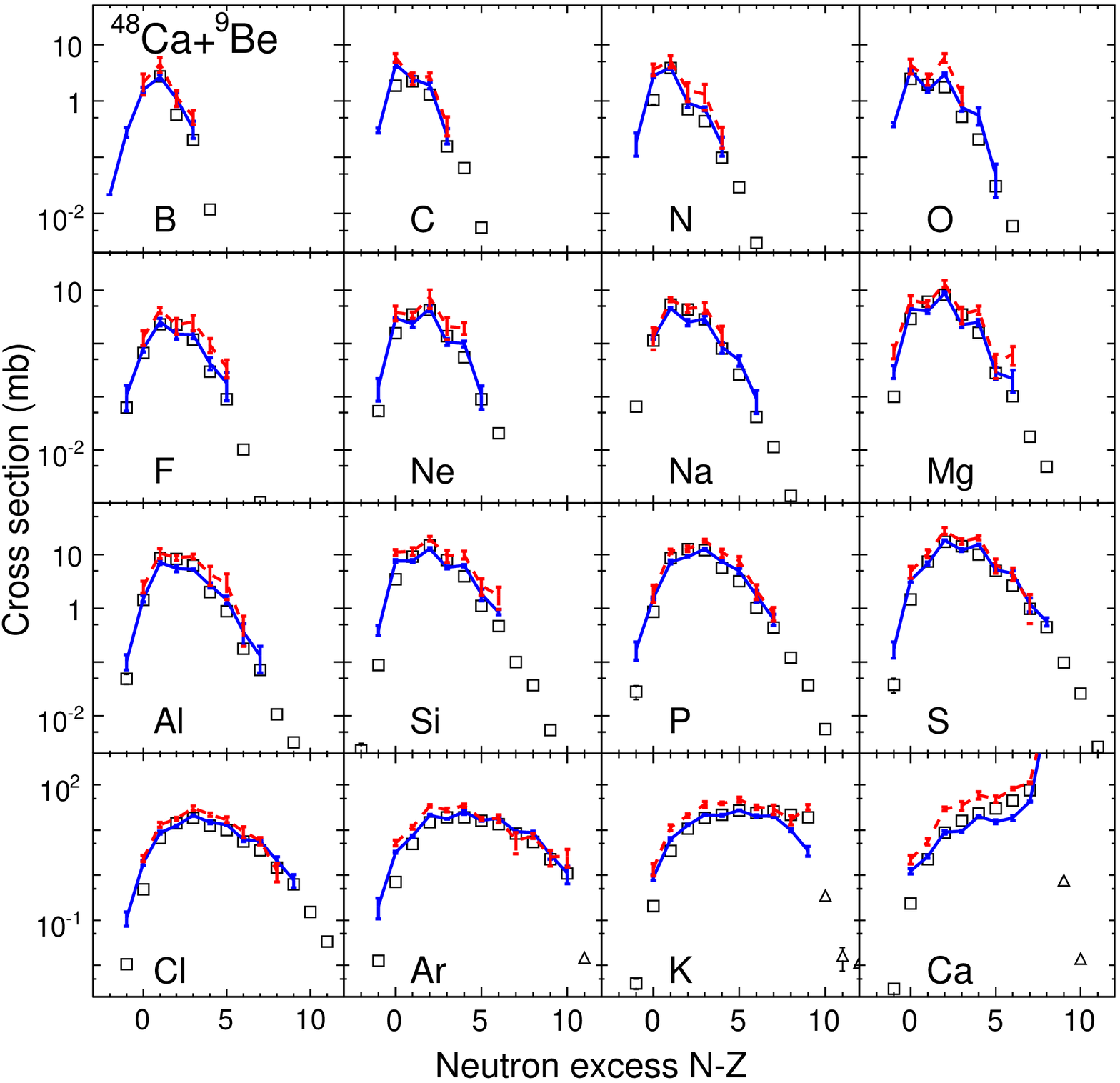}
\caption{(Color online) Fragmentation (open squares) and nucleon pick-up (open triangles) cross sections of $^{48}$Ca+$^9$Be reactions are compared to calculations by HIPSE (solid line) and AMD (dashed line) models. The error bars in the data are smaller than the symbols.}\label{fig10}
\end{center}
\end{figure}

\begin{figure}
\begin{center}
\includegraphics[width=0.8\textwidth]{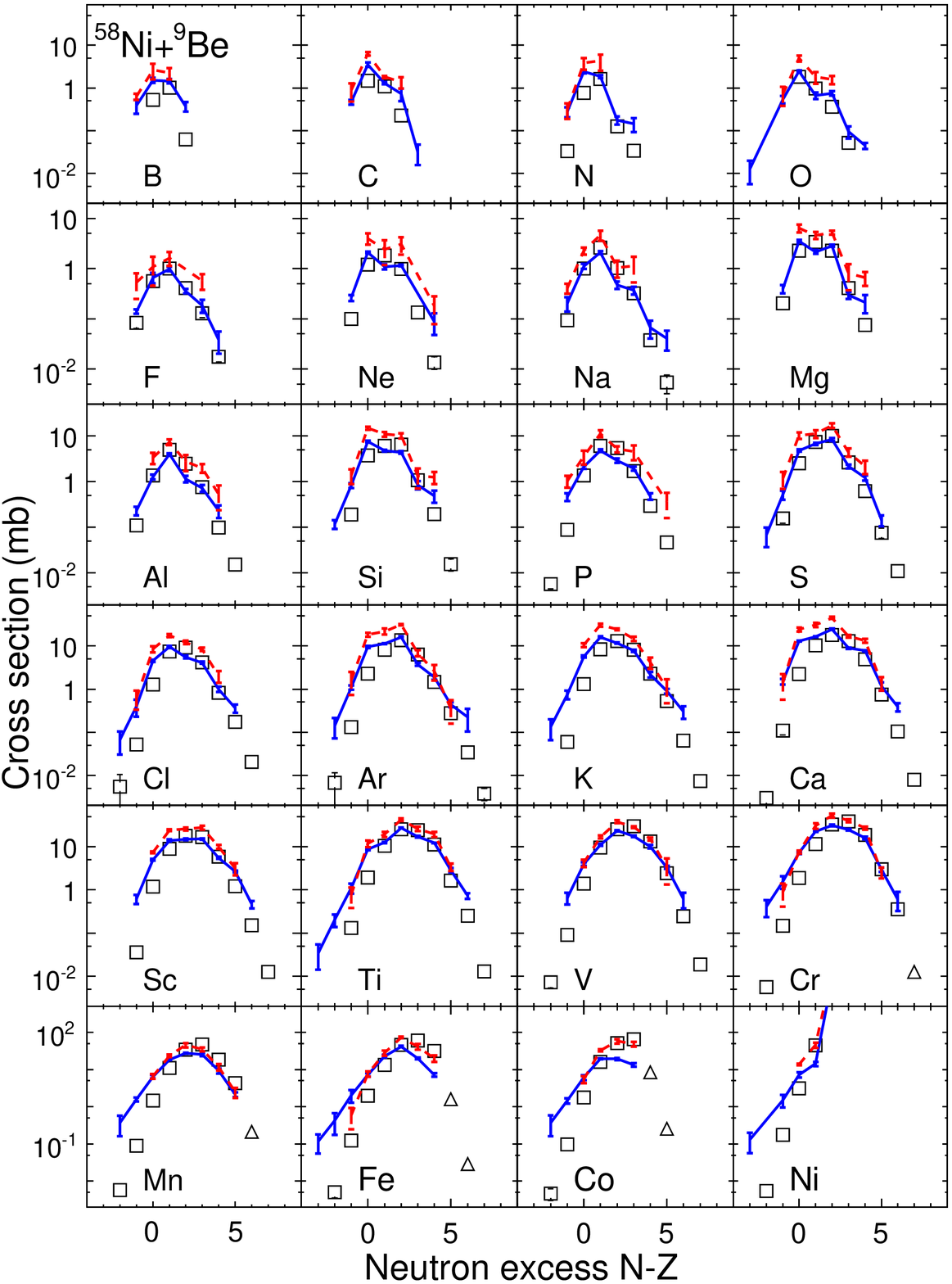}
\caption{(Color online) Fragmentation (open squares) and nucleon pick-up (open triangles) cross sections of $^{58}$Ni+$^9$Be reactions are compared to calculations by HIPSE (solid line) and AMD (dashed line) models. The error bars in the data are smaller than the symbols.}\label{fig11}
\end{center}
\end{figure}

\begin{figure}
\begin{center}
\includegraphics[width=.8\textwidth]{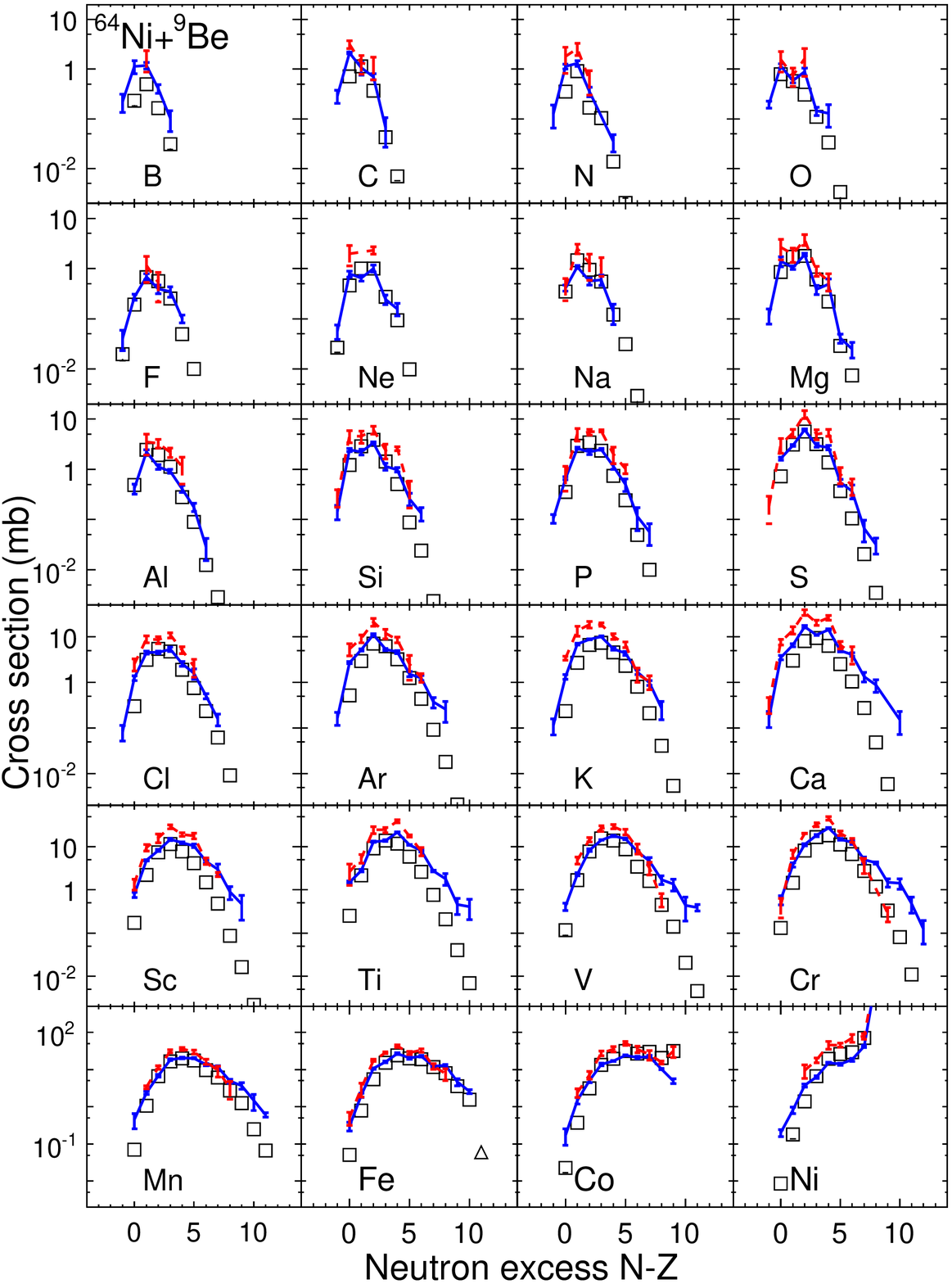}
\caption{(Color online) Fragmentation (open squares) and nucleon pick-up (open triangles) cross sections of $^{64}$Ni+$^9$Be reactions are compared to calculations by HIPSE (solid line) and AMD (dashed line) models. The error bars in the data are smaller than the symbols.}\label{fig12}
\end{center}
\end{figure}

Fig. \ref{fig9}--\ref{fig12} present comparisons of the measured isotopic cross sections without experimental efficiency corrections from the fragmentation of the $^{40}$Ca, $^{48}$Ca, $^{58}$Ni, and $^{64}$Ni primary beams on $^{9}$Be targets. They are presented in terms of isotopic distributions as a function of neutron excess, $N-Z$. Individual panels in Fig. \ref{fig9}--\ref{fig12} show isotopic distributions for different elements labeled with their respective chemical symbols. (Note that these experimental cross sections are different from those published in Ref. \cite{moc06,moc06a}. The previously published cross sections were corrected for angular and momentum transmission inefficiency based on parameterization without the knowledge of the collision dynamics.)

The HIPSE and AMD models are stochastic calculations. The lower limit of calculated cross sections depends on the number of simulated events, which were approximately 100,000 and 20,000 for each reaction for the HIPSE and AMD models, respectively. Overall the peaks of the isotope distributions are described well by both calculations for fragmentation of $^{40,48}$Ca and $^{58,64}$Ni beams. Except for the $^{40}$Ca+$^9$Be reaction, the AMD-predicted cross sections for the other three reactions are consistently higher than the data and the HIPSE model predictions. There are many input parameters and model details in the AMD model, which affect cross sections. For example, the cross sections may be sensitive to the transport-model input parameters such as the in-medium nucleon-nucleon cross sections. The nuclear structure and density profile of the projectile and target nuclei may also affect the fragment cross sections. More calculations with the AMD model will be needed to understand this discrepancy further. 

In the case of neutron-deficient projectiles such as $^{40}$Ca the HIPSE model predicts larger cross sections of the fragments with mass close to the projectile which in turn leads to over-estimations of the neutron-deficient isotopes.

\begin{figure}
\begin{center}
\includegraphics[width=0.9\textwidth]{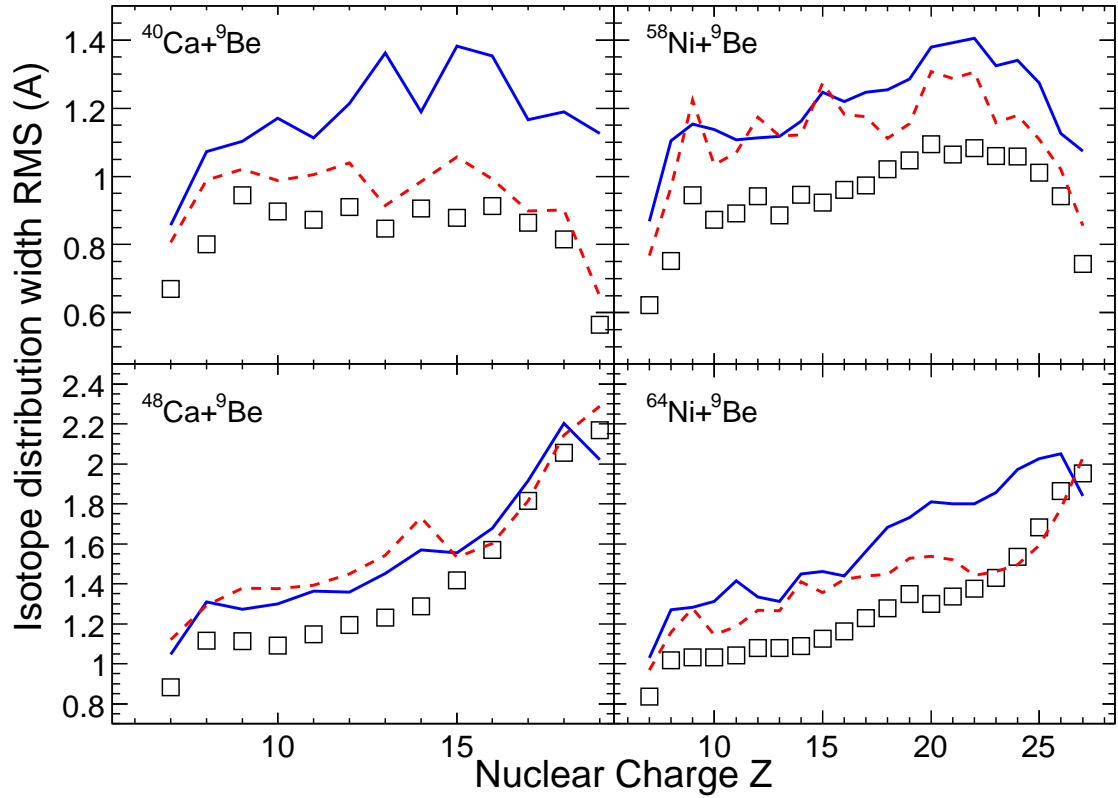}
\caption{(Color online) Widths of the isotope distributions shown in Fig.~\ref{fig9}--\ref{fig12}, expressed in terms of standard deviation (RMS) are plotted as a function of nuclear charge, $Z$, for all investigated reaction systems. Experimental data are shown with open squares, solid and dashed lines depict the HIPSE and AMD simulations, respectively.}\label{fig13}
\end{center}
\end{figure}

The isotope cross-section distributions calculated by the HIPSE model are generally wider than the experimental data for all investigated projectiles. Fig.~\ref{fig13} shows the RMS widths of the isotope distributions obtained from the data (open squares), HIPSE (solid line), and AMD (dashed line). The experimental widths are much better described by the AMD model. The wider isotopic widths predicted by the HIPSE model may be related to the larger fluctuations of the excitation energy of the primary fragments. Comparison of Fig.~\ref{fig2},\ref{fig3} and \ref{fig13} suggests that the isotopic widths are correlated to the mean excitation energy of the primary projectile-like residues produced in the models. The discrepancies in the isotopic widths are the largest for the $^{40}$Ca+$^9$Be reactions and least for the $^{48}$Ca+$^9$Be collisions. Similar discrepancies are observed in the mean excitation energy of the projectile-like particles.

%%%%%%%%%%%%%%%%%%%%%%%%%%%%%%%%%%%%%%%%%%%%%%%%%%%%%%
\section{Summary}\label{section:summary}
%%%%%%%%%%%%%%%%%%%%%%%%%%%%%%%%%%%%%%%%%%%%%%%%%%%%%%%%%%%%%%%%%%%%%%%%%%%%%%%%%%%%%%%%%%%%%%%%%%%%%%%%%%%%%%%%%%%%%%%%%%%%%%%%%%%%%%%
We carried out an extensive study of the projectile fragmentation reactions using the macroscopic-microscopic Heavy Ion Phase Space Model and the fully microscopic Antisymmetrized Molecular Dynamics Model. Even though these models were not developed to describe the projectile fragmentation process, the agreement between predictions and data is reasonable, especially when one considers that there is no effort to vary model parameters to fit the data and that these models were not developed to describe fragmentation reaction mechanisms.

These models go beyond the phenomenological models such as the EPAX or the Abrasion-Ablation model in describing the dynamics of the reactions as well as in predicting the excitation energy of the primary fragments. Both models give similar dependence of the excitation energy profiles as a function of removed nucleons from the projectile in reactions with Be targets. The saturation of the excitation energy is contradictory to the assumptions used in the AA models. The HIPSE model is able to reproduce the overall magnitude of the mass distributions, but the AMD model predictions are more consistent with the shape of the measured isotope distributions. The calculated final cross-section distributions are the results not only of the primary (fast) step of the nuclear reaction (modeled by HIPSE or AMD), but also of the secondary (slow) step, modeled by the statistical evaporation code (GEMINI). At 140 MeV/nucleon, the final fragment distributions are influenced significantly by the sequential decays. Hence, it is imperative to better understand the de-excitation part of the nuclear collision, the evaporation process, if we want to put the dynamical nuclear-collision calculations to a more stringent test. 

Due to the Monte-Carlo nature of the transport models, it is impractical to use these models to estimate the yield of the rare isotopes. However, all current models, including EPAX parameterization, cannot predict sufficiently accurately the yields of rare isotopes with extremely low cross sections. Dominance of the sequential decay processes suggests that parameterization based on statistical model considerations gives more accurate predictions of the yields of these rare nuclei \cite{moc07,tsa07}.

Finally, models such as AMD include information about transport mechanisms and their parameters. Of particular interest are asymmetry term of the nuclear equation of state, the in-medium nucleon-nucleon collisions, and cluster formation. Our analysis suggests that in addition to the cross section, other measured experimental quantities such as the momentum distributions may provide important constraints to these transport quantities.

%%%%%%%%%%%%%%%%%%%%%%%%%%%%%%%%%%%%%%%%%%%%%%%%%%%%%%%%%%%%%%%%%%%%%%%%%%%%%%%%%%%%%%%%%%%%%%%%%%%%%%%%%%%%%%%%%%%%%%%%%%%%%%%%%%%%%%%%
% Acknowledgement

% put your acknowledgments here.

\begin{acknowledgments}
A. Ono and D. Lacroix thank the National Superconducting Cyclotron Laboratory at Michigan State University for the support and hospitality during their sabbatical stays in 2005--2006 (A.O.) and 2006--2007 (D.L.). We would like to acknowledge the High Performance Computing Center \cite{hpc} at Michigan State University for facilitating CPU-intensive calculations with the AMD code. This work is supported by the National Science Foundation under Grant Nos. PHY-01-10253, PHY-0606007 and DE-FG02-04ER41313.
\end{acknowledgments}

%%%%%%%%%%%%%%%%%%%%%%%%%%%%%%%%%%%%%%%%%%%%%%%%%%%%%%%%%%%%%%%%%%%%%%%%%%%%%%%%%%%%%%%%%%%%%%%%%%%%%%%%%%%%%%%%%%%%%%%%%%%%%%%%%%%%%%%
%%% References::
%%%%%%%%%%%%%%%%%%%%%%%%%%%%%%%%%%%%%%%%%%%%%%%%%%%%%%%%%%%%%%%%%%%%%%%%%%%%%%%%%%%%%%%%%%%%%%%%%%%%%%%%%%%%%%%%%%%%%%%%%%%%%%%%%%%%%%%

%\bibliography{references}
%\bibliographystyle{unsrt}

\clearpage
%%%%%%%%%%%%%%%%%%%%%%%%%%%%%%%%%%%%%%%%%%%%%%%%%%%%%%%%%%%%%%%%%%%%%%%%%%%%%%%%%%%%%%%%%%%%%%%%%%%%%%%%%%%%%%%%%%%%%%%%%%%
%% Figures:

%%% cross section plots

\end{document}